\documentclass[%
superscriptaddress,
showpacs,preprintnumbers,
amsmath,amssymb,
aps,
prl,
linenumbers,
twocolumn
]{revtex4}
\usepackage{color}
\usepackage{graphicx}
\usepackage{dcolumn}
\usepackage{bm}

\begin {document}

\title{Origin of $1/f$ noise transition in hydration dynamics on a lipid membrane surface}

\author{Eiji Yamamoto}
\author{Takuma Akimoto}
\affiliation{%
  Department of Mechanical Engineering, Keio University, Yokohama, Japan
}%

\author{Masato Yasui}
\affiliation{%
  Department of Pharmacology, School of Medicine, Keio University, Shinjuku-ku, Tokyo, Japan
}%

\author{Kenji Yasuoka}
\affiliation{%
  Department of Mechanical Engineering, Keio University, Yokohama, Japan
}%



\begin{abstract}
Water molecules on lipid membrane surfaces are known to contribute to membrane stability by connecting lipid molecules and acting as a water bridge.
Although the number of water molecules near the membrane fluctuates dynamically, the hydration dynamics has been veiled.
Here we investigate residence statistics of water molecules on the surface of a lipid membrane using all-atom molecular dynamics simulations.
We show that hydration dynamics on the lipid membrane exhibit $1/f^\beta$ noise with two different power-law exponents, $\beta_l < 1$ and $\beta_h > 1$.
By constructing a dichotomous process for the hydration dynamics, we find that the process can be regarded as a non-Markov renewal process.
The result implies that the origin of the $1/f$ noise transition in hydration dynamics on the membrane surface is a combination of a power-law distribution with cutoff of interoccurrence times of switching events and a long-term correlation between the interoccurrence times.
\end{abstract}

\maketitle


In numerous natural systems, the power spectra $S(f)$ exhibit enigmatic $1/f$ noise:
\begin{equation}
S(f) \sim \frac{A}{f^\beta} ~ (0 < \beta < 2).
\end{equation}
at low frequencies.
In biological systems, $1/f$ noise has been reported for protein conformational dynamics~\cite{YangLuoKarnchanaphanurachLouieRechCovaXunXie2003,MinLuoCherayilKouXie2005,YamamotoAkimotoHiranoYasuiYasuoka2014}, DNA sequences~\cite{LiKaneko1992}, biorecognition~\cite{BizzarriCannistraro2013}, and ionic currents~\cite{BezrukovWinterhalter2000,MercikWeron2001,SiwyFuliifmmode2002,TasseritKoutsioubasLairezZalczerClochard2010}, implying that long-range correlated dynamics underlie biological processes.
Moreover, $1/f$ noise is involved in the regulation of permeation of water molecules in an aquaporin~\cite{YamamotoAkimotoHiranoYasuiYasuoka2014}.

There are many mathematical models that generate $1/f$ noise including stochastic models~\cite{MandelbrotVan1968,LowenTeich1993,Godr`echeLuck2001,DavidsenSchuster2002} and intermittent dynamical systems~\cite{Manneville1980,ProcacciaSchuster1983,Aizawa1984,GeiselNierwetbergZacherl1985}.
The power-law residence time distribution is one of the most thoroughly studied origins for $1/f$ noise~\cite{Manneville1980,ProcacciaSchuster1983,Aizawa1984,GeiselNierwetbergZacherl1985,Godr`echeLuck2001}. 
In dichotomous processes, the power spectrum shows $1/f$ noise when the distribution of residence times of each state follows a power-law distribution with divergent second moment.
For blinking quantum dots, which show a $1/f$ spectrum, residence times for ``on'' (bright) and ``off'' (dark) states have been experimentally shown to have a power-law distribution with a divergent mean~\cite{KunoFrommHamannGallagherNesbitt2000,BrokmannHermierMessinDesbiollesBouchaudDahan2003}.
In stochastic models, this divergent mean residence time violates the law of large numbers which causes the breakdown of ergodicity, non-stationarity, and aging~\cite{MargolinBarkai2006,HeBurovMetzlerBarkai2008,MiyaguchiAkimoto2011,NiemannKantzBarkai2013}.
Conversely, the divergent mean residence time implies an infinite invariant measure in dynamical systems~\cite{AkimotoAizawa2010} and that the time-averaged observables are intrinsically random~\cite{Akimoto2008,AkimotoAizawa2010}.

In our previous work, we found that the residence times of water molecules on the lipid membrane surfaces followed power-law distributions~\cite{YamamotoAkimotoHiranoYasuiYasuoka2013,YamamotoAkimotoYasuiYasuoka2014}.
Therefore, it is physically reasonable to expect that the hydration dynamics on membrane surfaces also obey $1/f$ noise.
Although little is known about the hydration dynamics, it is important to understand the dynamics of resident water molecules because these water molecules may play important roles in the overall dynamics of the membrane, and will affect membrane stability and biological reactions.
In fact, such water molecules stabilize the assembled lipid structures~\cite{Pasenkiewicz-GierulaTakaokaMiyagawaKitamuraKusumi1997,YamamotoAkimotoHiranoYasuiYasuoka2013}; this water retardation increases the efficiency of biological reactions~\cite{Ball2011,GrossmanBornHeydenTworowskiFieldsSagiHavenith2011,YamamotoAkimotoYasuiYasuoka2014}.
Water molecules enter and exit the hydration layer, and the number of water molecules near the lipid head group fluctuates.

In this letter, we perform a molecular dynamics (MD) simulation on water molecules plus a palmitoyl-oleoyl-phosphocholine (POPC) membrane at 310~K to investigate the hydration dynamics on the lipid surface [the details of the MD simulation are shown in~\cite{support}].
We find that fluctuations in the number of water molecules on the lipid surface show $1/f^\beta$ noise with two power-law exponents, {\it i.e.}, $\beta_l < 1$ at low frequencies and $\beta_h > 1$ at high frequencies, and that the residence time distributions for ``on'' and ``off'' states follow power-law distributions with exponential cutoffs.
Moreover, we construct a dichotomous process from the trajectory of the number of water molecules on a lipid molecule to clarify the origin of the two power-law exponents in the power spectrum.
By analyzing the constructed dichotomous process, we find that there is a long-term correlation in residence times, which causes two different power-law exponents in the power spectrum.

\begin{figure*}[tb]
\begin{center}
\includegraphics[width=135 mm,bb= 0 0 434 325]{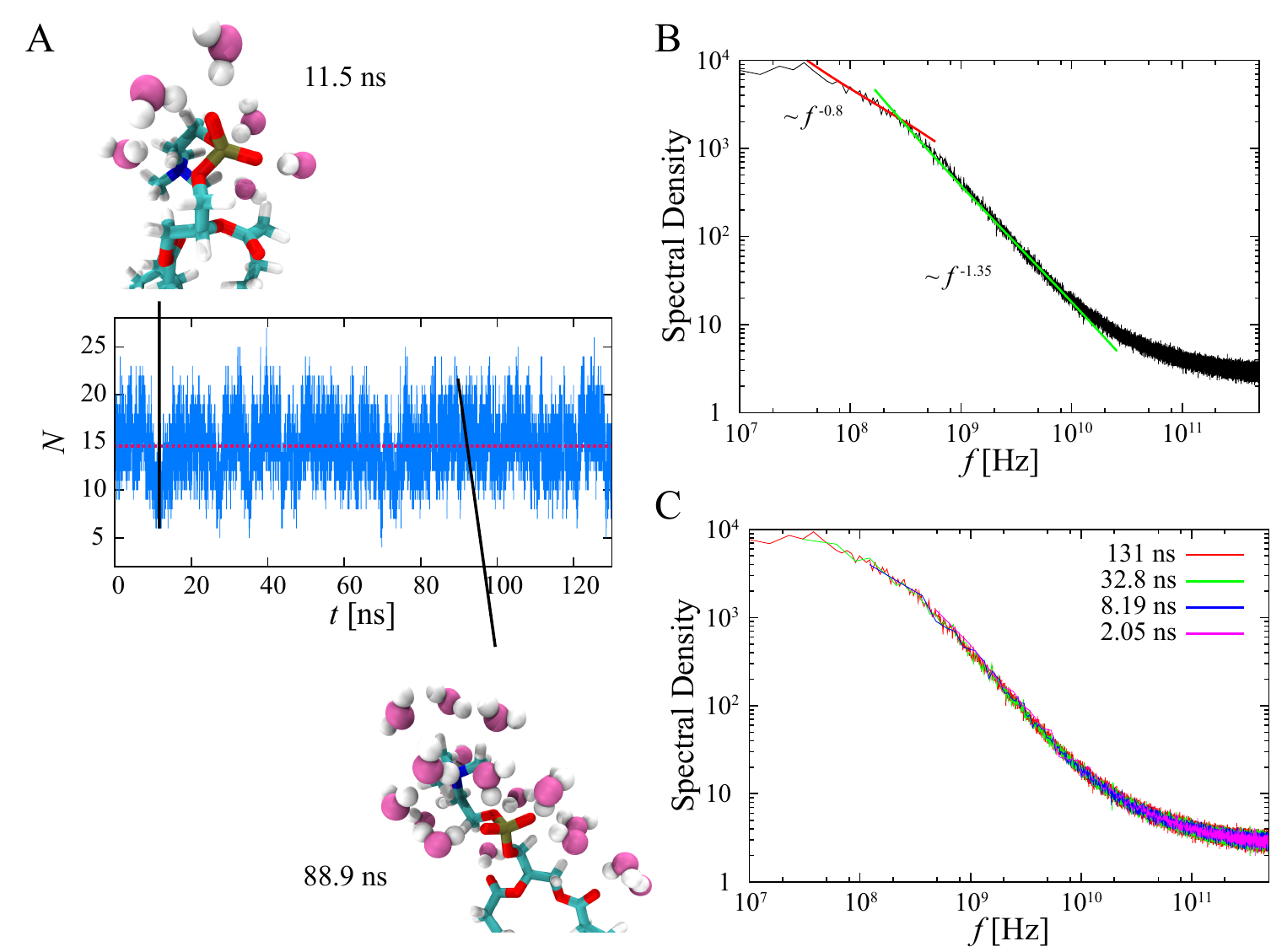}
\caption{Power spectrum of the number of water molecules.
(A)~Time series of number of water molecules on a lipid head group.
The red dashed line is the average number of water molecules on the lipid head group over this time period.
The outer windows show snapshots of water molecules around the lipid head group.
(B)~Ensemble-averaged PSD of number of water molecules.
We use 128 time series to obtain the ensemble-averaged PSD.
The solid lines represent power-law behavior for reference.
Total measurement time was 131~ns.
(C)~Ensemble-averaged PSD for four different measurement times: 2.05, 8.19, 32.8, and 131~ns.
The power spectra coincide without fitting.}
\end{center}
\end{figure*}

{\it Fluctuations of water molecules on the lipid head group.$-$}We recorded the number of water molecules for which the oxygens were within interatomic distances of 0.35~nm from all atoms in lipid head groups [Fig.~1A].
The number fluctuates around an average of about 14.
Figure~1B shows the ensemble-averaged power spectral density (PSD) obtained from the average of the power spectra for the number of water molecules at 128 lipid molecules.
The power spectrum exhibits two regimes with distinctive $1/f$ behavior.
For frequencies above a transition frequency $f_t$, we have $S(f) \propto f ^{-\beta_h}$ with $\beta_h = 1.35$, while below this frequency we have $S(f) \propto f ^{-\beta_l}$ with $\beta_l = 0.8$; furthermore, the PSD shows a plateau at low frequencies.
This crossover phenomenon is essential because $S(f) \propto f ^{-\beta}$ with $\beta \geq 1$ implies non-integrability and non-stationarity.
We have confirmed that $1/f$ fluctuations of the number of water molecules are observed in boxes and spheres near the membrane surfaces but not in bulk water.
A similar transition of the power-law exponent of the PSD has also been observed for the interchange dynamics of ``on'' and ``off'' states for quantum dot blinking~\cite{SadeghBarkaiKrapf2013}.
This behavior was described theoretically using an alternating renewal process, where the residence time distributions of ``on'' and ``off'' states are given by a power-law with an exponential cutoff $\psi_{\rm on}(\tau) \propto \tau^{-1-\alpha} e^{- \tau/\tau_{\rm on}}$ and a power-law $\psi_{\rm off}(\tau) \propto \tau^{-1-\alpha}$ where $\alpha < 1$, respectively~\cite{SadeghBarkaiKrapf2013}.
The transition frequency $f_t$ is related to the exponential cutoff in the quantum dot blinking experiment.
In this case, the PSD exhibits aging, non-stationarity, and weak ergodicity breaking because the ``off'' time does not have a finite mean.

To confirm whether the aging effect appears in the hydration dynamics on the lipid surface, we calculate the ensemble-averaged PSDs for different measurement times~[Fig.~1C].
The magnitudes of the PSDs do not depend on the measurement time $t$, {\it i.e.} there is no aging.
It follows that the power-law distribution with an exponential cutoff considered in~\cite{SadeghBarkaiKrapf2013} cannot explain hydration dynamics on lipid membranes.

\begin{figure}[tb]
\begin{center}
\includegraphics[width=75 mm,bb= 0 0 224 440]{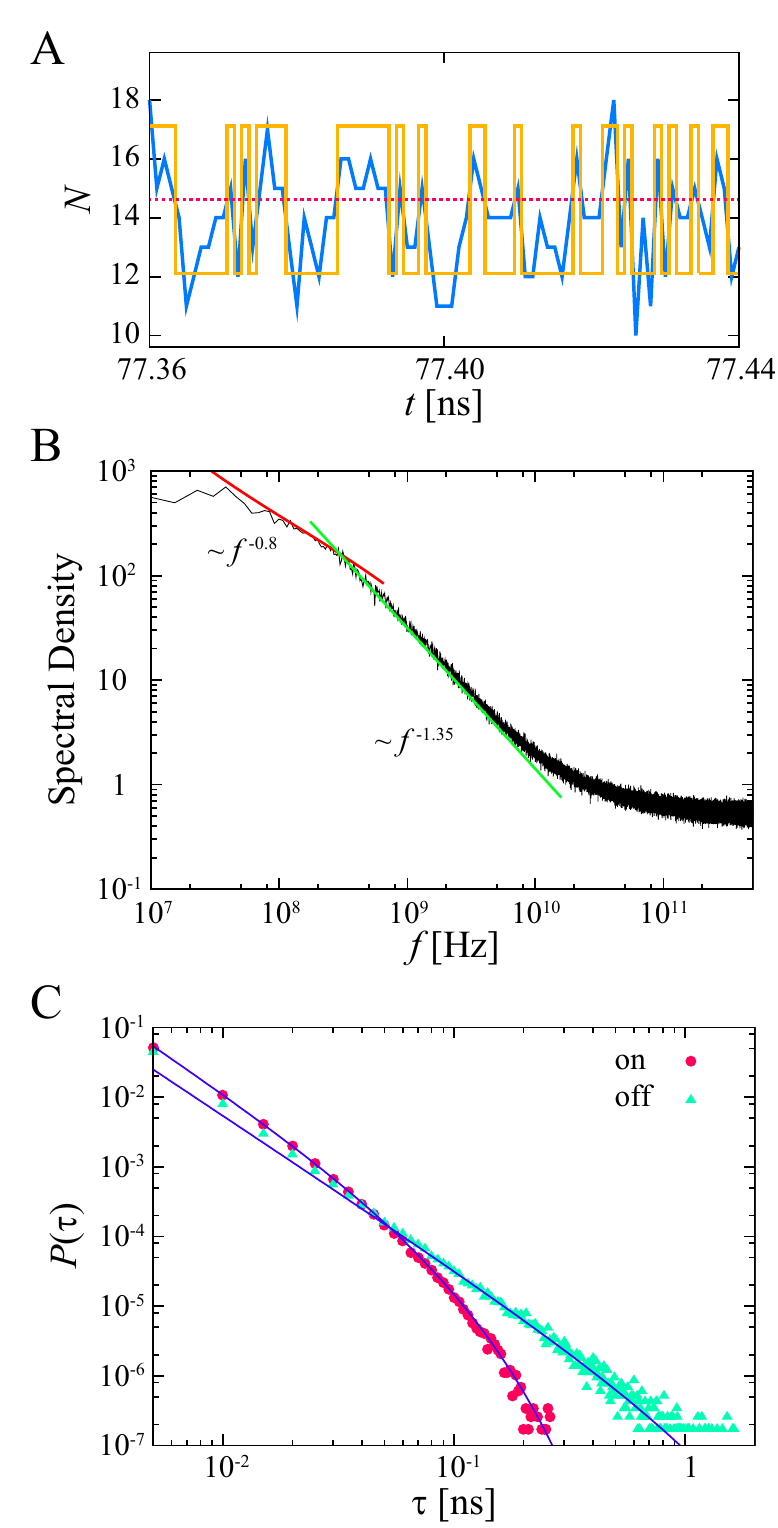}
\caption{$1/f$ noise in dichotomous processes.
(A)~Part of a time series of the number of water molecules on a lipid molecule (blue line); conversion of this data into ``on'' or ``off'' states (yellow line), depending on whether the number of water molecules is above or below the average (red dashed line).
(B)~Ensemble-averaged PSD of the time series of the two states.
The solid lines are shown as reference for higher and lower frequencies.
(C)~PDFs of residence times of ``on'' and ``off'' states.
Solid lines are fitted curves for power-law distributions with exponential cutoffs: $P(\tau)=A \tau ^{- (1+\alpha)} \exp (-\tau / \tau_c)$ ($\alpha = 1.2$, on: $\tau_c=59$, off: $\tau_c=1074$).}
\end{center}
\end{figure}

\begin{figure}[tb]
\begin{center}
\includegraphics[width=75 mm,bb= 0 0 205 152]{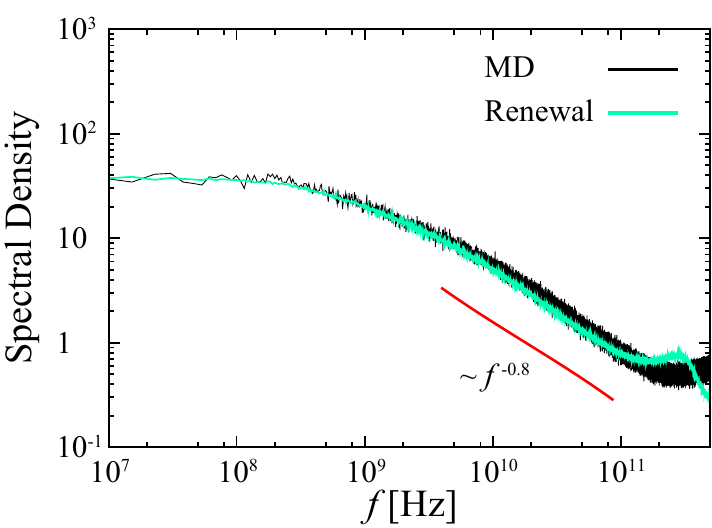}
\caption{Ensemble-averaged PSD of shuffles time series of the two states (black line).
Numerical simulation of Markov renewal process with two states; residence times are given by power-law distribution with exponential cutoff, where $\alpha = 1.2$, on: $\tau_c=60$, off: $\tau_c=1000$ (green line).
The solid line is shown for reference.}
\end{center}
\end{figure}

{\it Dichotomous process.$-$}To consider the origin of $1/f$ noise, we constructed a dichotomous, {\it i.e.} two state, process from the time series of the number of water molecules; the ``on'' ($N' = 1$) or ``off'' ($N' = -1$) states are when the number of water molecules on each lipid molecule is above or below, respectively, the average number [Fig.~2A].
Figure~2B shows the ensemble-averaged PSD for the time series of constructed dichotomous processes.
The obtained $1/f$ noise is the same as the ensemble-averaged PSD for the original time series [see Fig.~1B].
Figure~2C shows probability density functions (PDFs) of residence times for ``on'' and ``off'' states.
The PDFs follow power-law distributions with exponential cutoffs, $P(\tau)=A \tau ^{- (1 + \alpha)} \exp (-\tau / \tau_c)$, where the power-law exponent is $\alpha = 1.2$, and cutoffs for the PDFs of the ``on'' and ``off'' states are $\tau_c = 59$~ps and 1074~ps, respectively.
The plateau of the PSD at low frequencies comes from the exponential cutoffs in the power law distributions.

{\it Origin of the transit $1/f$ noise.$-$}One important question remains unclear: What is the origin of the transition in the $1/f$ noise?
In other words, does power law intermittency or long-term memory (as expected for a non-Markov process) contribute to the transition in the $1/f$ noise?
To address this question, we calculated the ensemble-averaged PSD for a shuffled time series of dichotomous processes, where residence times for ``on'' and ``off'' states were shuffled among themselves randomly.
The ensemble-averaged PSD of the shuffled time series is different from that of the original time series of the dichotomous process [Fig.~3].
The transition in the $1/f$ noise disappears, although the power spectrum shows $1/f$ noise at high frequencies even after shuffling.
The power-law exponent of $S(f) \propto f^{-\beta}$ at high frequencies is about $0.8$, and the PSD converges to a finite value at low frequencies.
This suggests that the transition in the $1/f$ noise originates from the non-Markovian nature of the hydration dynamics.
Following our observations, we performed a numerical simulation in which time series of ``on'' and ``off'' states were generated with random waiting times drawn from a power-law distribution with an exponential cutoff, where $\alpha = 1.2$, on: $\tau_c=60$, off: $\tau_c=1000$.
In Markovian dichotomous processes, the power-law exponent $\beta$ in the PSD is given by the power-law exponent in the residence time distribution, {\it i.e.}, $\beta = 2 - \alpha$ as $\alpha <2$~\cite{ProcacciaSchuster1983}.
The power-law exponent $\beta$ observed here in the PSD is consistent with this relationship.

To clarify the correlation of residence times, we considered three types of time series of residence times: $\{ \tau^{\rm on}_1, ..., \tau^{\rm on}_N  \}$, $\{ \tau^{\rm off}_1, ..., \tau^{\rm off}_N  \}$, and $\{ \tau^{\rm on}_1, \tau^{\rm off}_1, ..., \tau^{\rm on}_N, \tau^{\rm off}_N  \}$.
Figure~4A shows correlations between ``on'' and ``off'' residence times.
There are positive correlations of residence times between the previous ``on'' state and the current ``on'' state or the previous ``off'' state and the current ``off'' state, and negative correlations of residence times between an ``on'' state residence time and the next ``off'' state time or an ``off'' state residence time and the next ``on'' state time.
Moreover, the ensemble-averaged PSDs of the three types of time series of residence times exhibit $1/f$ noise [Fig.~4B].
This result means that the residence times have a long-term correlation.

What is a biological significance of $1/f$ noise in hydration dynamics on lipid membrane surfaces?
The roles played by the water molecules near the membrane depend upon their structure and dynamics.
The $1/f$ noise attributed to a non-Markov renewal process can contribute to the stability of the hydration layer, which is important for membrane stability and physiological processes.

\begin{figure}[tb]
\begin{center}
\includegraphics[width=75 mm,bb= 0 0 231 318]{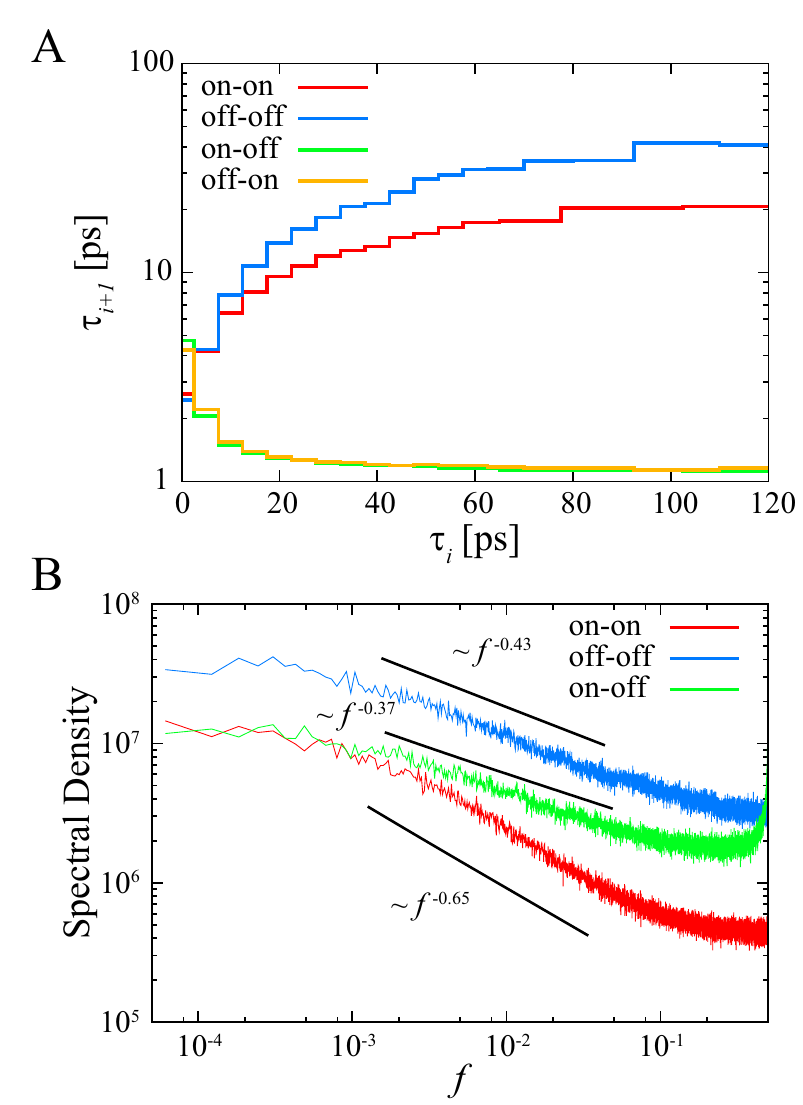}
\caption{Correlation of residence times.
(A)~Correlation of the residence times between $\tau_{i}$ and $\tau_{i+1}$.
Different color lines distinguish the pairs used for the analysis.
(B)~Ensemble-averaged PSD of residence times.
The solid lines are shown for reference.}
\end{center}
\end{figure}

In conclusion, we have used all-atom molecular dynamics simulations to show that the number of water molecules on the lipid molecules exhibits $1/f$ noise.
The power law exponents are different below and above the transition frequency $f_t$.
There is a transition from $\beta_l < 1$ at low frequencies to $\beta_h > 1$ at high frequencies, although the power spectrum does not break ergodicity.
Moreover, we provide evidence that the transition in the $1/f$ noise and ergodicity are caused by non-Markov power-law intermittency with exponential cutoff.
These results are relevant to a broad range of systems displaying $1/f$ fluctuations.

This work was supported by the Core Research for the Evolution Science and Technology (CREST) of the Japan Science and Technology Agency, JSPS KAKENHI (Grant-in-Aid for Challenging Exploratory Research) Grant No. 25630070, Keio University Program for the Advancement of Next Generation Research Projects, and MEXT Grant-in-Aid for the ``Program for Leading Graduate Schools''.


\begin{thebibliography}{32}
\expandafter\ifx\csname natexlab\endcsname\relax\def\natexlab#1{#1}\fi
\expandafter\ifx\csname bibnamefont\endcsname\relax
  \def\bibnamefont#1{#1}\fi
\expandafter\ifx\csname bibfnamefont\endcsname\relax
  \def\bibfnamefont#1{#1}\fi
\expandafter\ifx\csname citenamefont\endcsname\relax
  \def\citenamefont#1{#1}\fi
\expandafter\ifx\csname url\endcsname\relax
  \def\url#1{\texttt{#1}}\fi
\expandafter\ifx\csname urlprefix\endcsname\relax\fi
\providecommand{\bibinfo}[2]{#2}
\providecommand{\eprint}[2][]{\url{#2}}

\bibitem[{\citenamefont{Yang et~al.}(2003)\citenamefont{Yang, Luo,
  Karnchanaphanurach, Louie, Rech, Cova, Xun, and
  Xie}}]{YangLuoKarnchanaphanurachLouieRechCovaXunXie2003}
\bibinfo{author}{\bibfnamefont{H.}~\bibnamefont{Yang}},
  \bibinfo{author}{\bibfnamefont{G.}~\bibnamefont{Luo}},
  \bibinfo{author}{\bibfnamefont{P.}~\bibnamefont{Karnchanaphanurach}},
  \bibinfo{author}{\bibfnamefont{T.~M.} \bibnamefont{Louie}},
  \bibinfo{author}{\bibfnamefont{I.}~\bibnamefont{Rech}},
  \bibinfo{author}{\bibfnamefont{S.}~\bibnamefont{Cova}},
  \bibinfo{author}{\bibfnamefont{L.}~\bibnamefont{Xun}}, \bibnamefont{and}
  \bibinfo{author}{\bibfnamefont{X.~S.} \bibnamefont{Xie}},
  \bibinfo{journal}{Science} \textbf{\bibinfo{volume}{302}},
  \bibinfo{pages}{262} (\bibinfo{year}{2003}).

\bibitem[{\citenamefont{Min et~al.}(2005)\citenamefont{Min, Luo, Cherayil, Kou,
  and Xie}}]{MinLuoCherayilKouXie2005}
\bibinfo{author}{\bibfnamefont{W.}~\bibnamefont{Min}},
  \bibinfo{author}{\bibfnamefont{G.}~\bibnamefont{Luo}},
  \bibinfo{author}{\bibfnamefont{B.~J.} \bibnamefont{Cherayil}},
  \bibinfo{author}{\bibfnamefont{S.~C.} \bibnamefont{Kou}}, \bibnamefont{and}
  \bibinfo{author}{\bibfnamefont{X.~S.} \bibnamefont{Xie}},
  \bibinfo{journal}{Phys. Rev. Lett.} \textbf{\bibinfo{volume}{94}},
  \bibinfo{pages}{198302} (\bibinfo{year}{2005}).

\bibitem[{\citenamefont{Yamamoto
  et~al.}(2014{\natexlab{a}})\citenamefont{Yamamoto, Akimoto, Hirano, Yasui,
  and Yasuoka}}]{YamamotoAkimotoHiranoYasuiYasuoka2014}
\bibinfo{author}{\bibfnamefont{E.}~\bibnamefont{Yamamoto}},
  \bibinfo{author}{\bibfnamefont{T.}~\bibnamefont{Akimoto}},
  \bibinfo{author}{\bibfnamefont{Y.}~\bibnamefont{Hirano}},
  \bibinfo{author}{\bibfnamefont{M.}~\bibnamefont{Yasui}}, \bibnamefont{and}
  \bibinfo{author}{\bibfnamefont{K.}~\bibnamefont{Yasuoka}},
  \bibinfo{journal}{Phys. Rev. E} \textbf{\bibinfo{volume}{89}},
  \bibinfo{pages}{022718} (\bibinfo{year}{2014}{\natexlab{a}}).

\bibitem[{\citenamefont{Li and Kaneko}(1992)}]{LiKaneko1992}
\bibinfo{author}{\bibfnamefont{W.}~\bibnamefont{Li}} \bibnamefont{and}
  \bibinfo{author}{\bibfnamefont{K.}~\bibnamefont{Kaneko}},
  \bibinfo{journal}{Eumphys. Lett.} \textbf{\bibinfo{volume}{17}},
  \bibinfo{pages}{655} (\bibinfo{year}{1992}).

\bibitem[{\citenamefont{Bizzarri and
  Cannistraro}(2013)}]{BizzarriCannistraro2013}
\bibinfo{author}{\bibfnamefont{A.~R.} \bibnamefont{Bizzarri}} \bibnamefont{and}
  \bibinfo{author}{\bibfnamefont{S.}~\bibnamefont{Cannistraro}},
  \bibinfo{journal}{Phys. Rev. Lett.} \textbf{\bibinfo{volume}{110}},
  \bibinfo{pages}{048104} (\bibinfo{year}{2013}).

\bibitem[{\citenamefont{Bezrukov and
  Winterhalter}(2000)}]{BezrukovWinterhalter2000}
\bibinfo{author}{\bibfnamefont{S.~M.} \bibnamefont{Bezrukov}} \bibnamefont{and}
  \bibinfo{author}{\bibfnamefont{M.}~\bibnamefont{Winterhalter}},
  \bibinfo{journal}{Phys. Rev. Lett.} \textbf{\bibinfo{volume}{85}},
  \bibinfo{pages}{202} (\bibinfo{year}{2000}).

\bibitem[{\citenamefont{Mercik and Weron}(2001)}]{MercikWeron2001}
\bibinfo{author}{\bibfnamefont{S.}~\bibnamefont{Mercik}} \bibnamefont{and}
  \bibinfo{author}{\bibfnamefont{K.}~\bibnamefont{Weron}},
  \bibinfo{journal}{Phys. Rev. E} \textbf{\bibinfo{volume}{63}},
  \bibinfo{pages}{051910} (\bibinfo{year}{2001}).

\bibitem[{\citenamefont{Siwy and Fuli\ifmmode~\acute{n}\else
  \'{n}\fi{}ski}(2002)}]{SiwyFuliifmmode2002}
\bibinfo{author}{\bibfnamefont{Z.}~\bibnamefont{Siwy}} \bibnamefont{and}
  \bibinfo{author}{\bibfnamefont{A.}~\bibnamefont{Fuli\ifmmode~\acute{n}\else
  \'{n}\fi{}ski}}, \bibinfo{journal}{Phys. Rev. Lett.}
  \textbf{\bibinfo{volume}{89}}, \bibinfo{pages}{158101}
  (\bibinfo{year}{2002}).

\bibitem[{\citenamefont{Tasserit et~al.}(2010)\citenamefont{Tasserit,
  Koutsioubas, Lairez, Zalczer, and
  Clochard}}]{TasseritKoutsioubasLairezZalczerClochard2010}
\bibinfo{author}{\bibfnamefont{C.}~\bibnamefont{Tasserit}},
  \bibinfo{author}{\bibfnamefont{A.}~\bibnamefont{Koutsioubas}},
  \bibinfo{author}{\bibfnamefont{D.}~\bibnamefont{Lairez}},
  \bibinfo{author}{\bibfnamefont{G.}~\bibnamefont{Zalczer}}, \bibnamefont{and}
  \bibinfo{author}{\bibfnamefont{M.-C.} \bibnamefont{Clochard}},
  \bibinfo{journal}{Phys. Rev. Lett.} \textbf{\bibinfo{volume}{105}},
  \bibinfo{pages}{260602} (\bibinfo{year}{2010}).

\bibitem[{\citenamefont{Lowen and Teich}(1993)}]{LowenTeich1993}
\bibinfo{author}{\bibfnamefont{S.~B.} \bibnamefont{Lowen}} \bibnamefont{and}
  \bibinfo{author}{\bibfnamefont{M.~C.} \bibnamefont{Teich}},
  \bibinfo{journal}{Phys. Rev. E} \textbf{\bibinfo{volume}{47}},
  \bibinfo{pages}{992} (\bibinfo{year}{1993}).

\bibitem[{\citenamefont{Davidsen and Schuster}(2002)}]{DavidsenSchuster2002}
\bibinfo{author}{\bibfnamefont{J.}~\bibnamefont{Davidsen}} \bibnamefont{and}
  \bibinfo{author}{\bibfnamefont{H.~G.} \bibnamefont{Schuster}},
  \bibinfo{journal}{Phys. Rev. E} \textbf{\bibinfo{volume}{65}},
  \bibinfo{pages}{026120} (\bibinfo{year}{2002}).

\bibitem[{\citenamefont{Godr{\`e}che and Luck}(2001)}]{Godr`echeLuck2001}
\bibinfo{author}{\bibfnamefont{C.}~\bibnamefont{Godr{\`e}che}}
  \bibnamefont{and} \bibinfo{author}{\bibfnamefont{J.~M.} \bibnamefont{Luck}},
  \bibinfo{journal}{J. Stat. Phys.} \textbf{\bibinfo{volume}{104}},
  \bibinfo{pages}{489} (\bibinfo{year}{2001}).

\bibitem[{\citenamefont{Mandelbrot and Van~Ness}(1968)}]{MandelbrotVan1968}
\bibinfo{author}{\bibfnamefont{B.~B.} \bibnamefont{Mandelbrot}}
  \bibnamefont{and} \bibinfo{author}{\bibfnamefont{J.~W.}
  \bibnamefont{Van~Ness}}, \bibinfo{journal}{SIAM Rev.}
  \textbf{\bibinfo{volume}{10}}, \bibinfo{pages}{422} (\bibinfo{year}{1968}).

\bibitem[{\citenamefont{Manneville}(1980)}]{Manneville1980}
\bibinfo{author}{\bibfnamefont{P.}~\bibnamefont{Manneville}},
  \bibinfo{journal}{J. Physique} \textbf{\bibinfo{volume}{41}},
  \bibinfo{pages}{1235} (\bibinfo{year}{1980}).

\bibitem[{\citenamefont{Procaccia and Schuster}(1983)}]{ProcacciaSchuster1983}
\bibinfo{author}{\bibfnamefont{I.}~\bibnamefont{Procaccia}} \bibnamefont{and}
  \bibinfo{author}{\bibfnamefont{H.}~\bibnamefont{Schuster}},
  \bibinfo{journal}{Phys. Rev. A} \textbf{\bibinfo{volume}{28}},
  \bibinfo{pages}{1210} (\bibinfo{year}{1983}).

\bibitem[{\citenamefont{Aizawa}(1984)}]{Aizawa1984}
\bibinfo{author}{\bibfnamefont{Y.}~\bibnamefont{Aizawa}},
  \bibinfo{journal}{Prog. Theor. Phys.} \textbf{\bibinfo{volume}{72}},
  \bibinfo{pages}{659} (\bibinfo{year}{1984}).

\bibitem[{\citenamefont{Geisel et~al.}(1985)\citenamefont{Geisel, Nierwetberg,
  and Zacherl}}]{GeiselNierwetbergZacherl1985}
\bibinfo{author}{\bibfnamefont{T.}~\bibnamefont{Geisel}},
  \bibinfo{author}{\bibfnamefont{J.}~\bibnamefont{Nierwetberg}},
  \bibnamefont{and} \bibinfo{author}{\bibfnamefont{A.}~\bibnamefont{Zacherl}},
  \bibinfo{journal}{Phys. Rev. Lett.} \textbf{\bibinfo{volume}{54}},
  \bibinfo{pages}{616} (\bibinfo{year}{1985}).

\bibitem[{\citenamefont{Kuno et~al.}(2000)\citenamefont{Kuno, Fromm, Hamann,
  Gallagher, and Nesbitt}}]{KunoFrommHamannGallagherNesbitt2000}
\bibinfo{author}{\bibfnamefont{M.}~\bibnamefont{Kuno}},
  \bibinfo{author}{\bibfnamefont{D.~P.} \bibnamefont{Fromm}},
  \bibinfo{author}{\bibfnamefont{H.~F.} \bibnamefont{Hamann}},
  \bibinfo{author}{\bibfnamefont{A.}~\bibnamefont{Gallagher}},
  \bibnamefont{and} \bibinfo{author}{\bibfnamefont{D.~J.}
  \bibnamefont{Nesbitt}}, \bibinfo{journal}{J. Chem. Phys.}
  \textbf{\bibinfo{volume}{112}}, \bibinfo{pages}{3117} (\bibinfo{year}{2000}).

\bibitem[{\citenamefont{Brokmann et~al.}(2003)\citenamefont{Brokmann, Hermier,
  Messin, Desbiolles, Bouchaud, and
  Dahan}}]{BrokmannHermierMessinDesbiollesBouchaudDahan2003}
\bibinfo{author}{\bibfnamefont{X.}~\bibnamefont{Brokmann}},
  \bibinfo{author}{\bibfnamefont{J.-P.} \bibnamefont{Hermier}},
  \bibinfo{author}{\bibfnamefont{G.}~\bibnamefont{Messin}},
  \bibinfo{author}{\bibfnamefont{P.}~\bibnamefont{Desbiolles}},
  \bibinfo{author}{\bibfnamefont{J.-P.} \bibnamefont{Bouchaud}},
  \bibnamefont{and} \bibinfo{author}{\bibfnamefont{M.}~\bibnamefont{Dahan}},
  \bibinfo{journal}{Phys. Rev. Lett.} \textbf{\bibinfo{volume}{90}},
  \bibinfo{pages}{120601} (\bibinfo{year}{2003}).

\bibitem[{\citenamefont{Margolin and Barkai}(2006)}]{MargolinBarkai2006}
\bibinfo{author}{\bibfnamefont{G.}~\bibnamefont{Margolin}} \bibnamefont{and}
  \bibinfo{author}{\bibfnamefont{E.}~\bibnamefont{Barkai}},
  \bibinfo{journal}{J. Stat. Phys.} \textbf{\bibinfo{volume}{122}},
  \bibinfo{pages}{137} (\bibinfo{year}{2006}).

\bibitem[{\citenamefont{He et~al.}(2008)\citenamefont{He, Burov, Metzler, and
  Barkai}}]{HeBurovMetzlerBarkai2008}
\bibinfo{author}{\bibfnamefont{Y.}~\bibnamefont{He}},
  \bibinfo{author}{\bibfnamefont{S.}~\bibnamefont{Burov}},
  \bibinfo{author}{\bibfnamefont{R.}~\bibnamefont{Metzler}}, \bibnamefont{and}
  \bibinfo{author}{\bibfnamefont{E.}~\bibnamefont{Barkai}},
  \bibinfo{journal}{Phys. Rev. Lett.} \textbf{\bibinfo{volume}{101}},
  \bibinfo{pages}{058101} (\bibinfo{year}{2008}).

\bibitem[{\citenamefont{Miyaguchi and Akimoto}(2011)}]{MiyaguchiAkimoto2011}
\bibinfo{author}{\bibfnamefont{T.}~\bibnamefont{Miyaguchi}} \bibnamefont{and}
  \bibinfo{author}{\bibfnamefont{T.}~\bibnamefont{Akimoto}},
  \bibinfo{journal}{Phys. Rev. E} \textbf{\bibinfo{volume}{83}},
  \bibinfo{pages}{062101} (\bibinfo{year}{2011}).

\bibitem[{\citenamefont{Niemann et~al.}(2013)\citenamefont{Niemann, Kantz, and
  Barkai}}]{NiemannKantzBarkai2013}
\bibinfo{author}{\bibfnamefont{M.}~\bibnamefont{Niemann}},
  \bibinfo{author}{\bibfnamefont{H.}~\bibnamefont{Kantz}}, \bibnamefont{and}
  \bibinfo{author}{\bibfnamefont{E.}~\bibnamefont{Barkai}},
  \bibinfo{journal}{Phys. Rev. Lett.} \textbf{\bibinfo{volume}{110}},
  \bibinfo{pages}{140603} (\bibinfo{year}{2013}).

\bibitem[{\citenamefont{Akimoto and Aizawa}(2010)}]{AkimotoAizawa2010}
\bibinfo{author}{\bibfnamefont{T.}~\bibnamefont{Akimoto}} \bibnamefont{and}
  \bibinfo{author}{\bibfnamefont{Y.}~\bibnamefont{Aizawa}},
  \bibinfo{journal}{Chaos} \textbf{\bibinfo{volume}{20}},
  \bibinfo{pages}{033110} (\bibinfo{year}{2010}).

\bibitem[{\citenamefont{Akimoto}(2008)}]{Akimoto2008}
\bibinfo{author}{\bibfnamefont{T.}~\bibnamefont{Akimoto}}, \bibinfo{journal}{J.
  Stat. Phys.} \textbf{\bibinfo{volume}{132}}, \bibinfo{pages}{171}
  (\bibinfo{year}{2008}).

\bibitem[{\citenamefont{Yamamoto et~al.}(2013)\citenamefont{Yamamoto, Akimoto,
  Hirano, Yasui, and Yasuoka}}]{YamamotoAkimotoHiranoYasuiYasuoka2013}
\bibinfo{author}{\bibfnamefont{E.}~\bibnamefont{Yamamoto}},
  \bibinfo{author}{\bibfnamefont{T.}~\bibnamefont{Akimoto}},
  \bibinfo{author}{\bibfnamefont{Y.}~\bibnamefont{Hirano}},
  \bibinfo{author}{\bibfnamefont{M.}~\bibnamefont{Yasui}}, \bibnamefont{and}
  \bibinfo{author}{\bibfnamefont{K.}~\bibnamefont{Yasuoka}},
  \bibinfo{journal}{Phys. Rev. E} \textbf{\bibinfo{volume}{87}},
  \bibinfo{pages}{052715} (\bibinfo{year}{2013}).

\bibitem[{\citenamefont{Yamamoto
  et~al.}(2014{\natexlab{b}})\citenamefont{Yamamoto, Akimoto, Yasui, and
  Yasuoka}}]{YamamotoAkimotoYasuiYasuoka2014}
\bibinfo{author}{\bibfnamefont{E.}~\bibnamefont{Yamamoto}},
  \bibinfo{author}{\bibfnamefont{T.}~\bibnamefont{Akimoto}},
  \bibinfo{author}{\bibfnamefont{M.}~\bibnamefont{Yasui}}, \bibnamefont{and}
  \bibinfo{author}{\bibfnamefont{K.}~\bibnamefont{Yasuoka}},
  \bibinfo{journal}{Sci. Rep.} \textbf{\bibinfo{volume}{4}},
  \bibinfo{pages}{4720} (\bibinfo{year}{2014}{\natexlab{b}}).

\bibitem[{\citenamefont{Pasenkiewicz-Gierula
  et~al.}(1997)\citenamefont{Pasenkiewicz-Gierula, Takaoka, Miyagawa, Kitamura,
  and Kusumi}}]{Pasenkiewicz-GierulaTakaokaMiyagawaKitamuraKusumi1997}
\bibinfo{author}{\bibfnamefont{M.}~\bibnamefont{Pasenkiewicz-Gierula}},
  \bibinfo{author}{\bibfnamefont{Y.}~\bibnamefont{Takaoka}},
  \bibinfo{author}{\bibfnamefont{H.}~\bibnamefont{Miyagawa}},
  \bibinfo{author}{\bibfnamefont{K.}~\bibnamefont{Kitamura}}, \bibnamefont{and}
  \bibinfo{author}{\bibfnamefont{A.}~\bibnamefont{Kusumi}},
  \bibinfo{journal}{J. Phys. Chem. A} \textbf{\bibinfo{volume}{101}},
  \bibinfo{pages}{3677} (\bibinfo{year}{1997}).

\bibitem[{\citenamefont{Ball}(2011)}]{Ball2011}
\bibinfo{author}{\bibfnamefont{P.}~\bibnamefont{Ball}},
  \bibinfo{journal}{Nature} \textbf{\bibinfo{volume}{478}},
  \bibinfo{pages}{467} (\bibinfo{year}{2011}).

\bibitem[{\citenamefont{Grossman et~al.}(2011)\citenamefont{Grossman, Born,
  Heyden, Tworowski, Fields, Sagi, and
  Havenith}}]{GrossmanBornHeydenTworowskiFieldsSagiHavenith2011}
\bibinfo{author}{\bibfnamefont{M.}~\bibnamefont{Grossman}},
  \bibinfo{author}{\bibfnamefont{B.}~\bibnamefont{Born}},
  \bibinfo{author}{\bibfnamefont{M.}~\bibnamefont{Heyden}},
  \bibinfo{author}{\bibfnamefont{D.}~\bibnamefont{Tworowski}},
  \bibinfo{author}{\bibfnamefont{G.~B.} \bibnamefont{Fields}},
  \bibinfo{author}{\bibfnamefont{I.}~\bibnamefont{Sagi}}, \bibnamefont{and}
  \bibinfo{author}{\bibfnamefont{M.}~\bibnamefont{Havenith}},
  \bibinfo{journal}{Nat. Struct. Mol. Biol.} \textbf{\bibinfo{volume}{18}},
  \bibinfo{pages}{1102} (\bibinfo{year}{2011}).

\bibitem[{\citenamefont{{See Supplemental Material for additional figures and
  mathematical analyses.}}()}]{support}
\bibinfo{author}{\bibnamefont{{See supplemental material for the details of the MD simulation.}}}

\bibitem[{\citenamefont{Sadegh et~al.}(2013)\citenamefont{Sadegh, Barkai, and
  Krapf}}]{SadeghBarkaiKrapf2013}
\bibinfo{author}{\bibfnamefont{S.}~\bibnamefont{Sadegh}},
  \bibinfo{author}{\bibfnamefont{E.}~\bibnamefont{Barkai}}, \bibnamefont{and}
  \bibinfo{author}{\bibfnamefont{D.}~\bibnamefont{Krapf}},
  \bibinfo{journal}{arXiv:1312.3561}  (\bibinfo{year}{2013}).

\end{thebibliography}

\end{document}